\documentclass{pasa}

\title[Black Hole Mass Estimation]{Black Hole Mass Estimation: How good is the virial estimate?}
\author[S.~Yong et al.]{Suk Yee Yong\thanks{E-mail: \texttt{syong1@student.unimelb.edu.au}}, Rachel L.~Webster \and Anthea L.~King\\
\affil{School of Physics, University of Melbourne, Parkville, VIC 3010, Australia}}%
\jid{PASA}
\doi{10.1017/pas.\the\year.xxx}
\jyear{\the\year}

\usepackage{amsmath,amssymb,gensymb}
\usepackage{graphicx, xcolor}
\usepackage[authoryear]{natbib}
\bibpunct{(}{)}{;}{a}{}{,}
\defcitealias{Murray+:1995}{M95}
\defcitealias{Elvis:2004}{E04}
\newcommand{\subtext}[2]{\ensuremath{#1_{\text{#2}}}} 
\newcommand{\ion}[2]{#1\,{\scshape{#2}}} 
\newcommand{\ergs}{\ifmmode {\rm erg\,s}^{-1} \else erg\,s$^{-1}$\fi} 
\newcommand{\kms}{\ifmmode {\rm km\,s}^{-1} \else km\,s$^{-1}$\fi} 
\newcommand{\diff}{\mathop{}\!\mathrm{d}} 

\begin{document}

\begin{abstract}
Black hole mass is a key factor in determining how a black hole interacts with its environment. However, the determination of black hole masses at high redshifts depends on secondary mass estimators, which are based on empirical relationships and broad approximations. A dynamical disk wind broad line region (BLR) model of active galactic nuclei (AGN) is built in order to test the impact on the black hole mass calculation due to different BLR geometries and the inclination of the AGN. Monte Carlo simulations of two disk wind models are constructed to recover the virial scale factor, $f$, at various inclination angles. The resulting $f$ values strongly correlate with inclination angle, with large $f$ values associated with small inclination angles (close to face-on) and small $f$ values with large inclination angles (close to edge-on).

The $f$ factors are consistent with previously determined $f$ values, found from empirical relationships. Setting $f$ as a constant may introduce a bias into virial black hole mass estimates for a large sample of AGN. However, the extent of the bias depends on the line width characterisation (e.g.\ full width at half maximum (FWHM) or line dispersion). Masses estimated using \subtext{f}{FWHM} tend to biased towards larger masses, but this can be corrected by calibrating for the width or shape of the emission line.
\end{abstract}


\maketitle

\section{Introduction} \label{sec:intro}

Black holes are widely believed to be located at the centre of most galaxies, both active and quiescent galaxies \citep{Kormendy+Richstone:1995,Richstone:1995,Ferrarese+Ford:2005,Kormendy+Ho:2013}. Relationships have been observed between the mass of the black hole and the properties of the host-galaxy, namely stellar velocity dispersion \citep[the $\subtext{M}{BH} \text{--} \sigma_{*}$ relation;][]{Ferrarese+Merritt:2000,Gultekin+:2009,McConnell+Ma:2013}, light concentration \citep[the $\subtext{M}{BH} \text{--} C_{r_{c}}$ relation;][]{Graham+:2001}, bulge luminosity and bulge stellar mass \citep[the $\subtext{M}{BH} \text{--} \subtext{L}{bulge}$ and $\subtext{M}{BH} \text{--} \subtext{M}{bulge}$ relations;][]{Magorrian+:1998,Marconi+Hunt:2003,McConnell+Ma:2013}. The origin of these correlations and the role of the central black hole in galaxy evolution, are still not well understood \citep{Silk+Rees:1998,King:2003,King:2005,DiMatteo+:2005,Murray+:2005,DiMatteo+:2008,Park+:2015}. The properties of a black hole can be related to its mass, \subtext{M}{BH}, and to understand the interplay between the black hole and its host galaxy, we require precise and accurate \subtext{M}{BH} measurements over a broad range of galaxy properties and cosmic time.

The value of \subtext{M}{BH} can be measured directly using the dynamics of stars or gas in close proximity to the black hole \citep{Ferrarese+Ford:2005,McConnell+Ma:2013}. However, this method is limited to the local universe due to the high spatial resolution required. An alternative method of black hole mass estimation is reverberation mapping \citep[RM;][]{Blandford+McKee:1982,Peterson:1993} of active galactic nuclei (AGN). Variable continuum emission originating from the accretion disk is absorbed by nearby gas deep within the gravitational potential of the black hole (broad line region; BLR). The BLR gas reprocesses this radiation and emits Doppler broadened emission-lines. The corresponding emission-line flux is observed to vary in response to the continuum flux in a roughly linear fashion with a time delay, $\tau$. This time delay corresponds to the light travel time to the mean responsivity weighted distance to the BLR from the accretion disk. RM is based on the assumption that there is a simple, though not necessarily linear, relationship between the observed continuum and the ionising continuum \citep{Peterson:1993}. In general, the emission line response has been found to be approximately linear in fashion; however, non-linear responses have been observed in NGC7469 \citep{Peterson+:2014}, NGC5548 \citep{DeRosa+:2015} during the second half of the campaign, and J080131 \citep{Du+:2015}. The mechanism causing the observed non-linear response in these objects is not well understood.

Under the assumption that the gas in the BLR is virialised and its motion is dominated by the gravitational field of the central black hole, the mass of the black hole is \citep{Peterson+Wandel:1999}
\begin{equation}
\subtext{M}{BH}=f\left(\frac{\Delta V^{2}R}{G}\right)=f\subtext{M}{vir},
\label{eqn:mbh}
\end{equation}
where $R=c\tau$ is the radius of the emitting line, $c$ is the speed of light, and $G$ is the gravitational constant. The velocity dispersion, denoted by $\Delta V$, is determined from the width of an individual broad emission line by measuring the full width at half maximum (FWHM) or the line dispersion, \subtext{\sigma}{line}, and $f$ is the virial factor that links the line-of-sight virial product, \subtext{M}{vir}, to the true black hole mass.

Reverberation mapping has yielded masses for approximately 60 AGN \citep{Bentz+Katz:2015} and the values of $R$ have been found to exhibit a tight power law relationship with the AGN continuum luminosity, $\lambda L_{\lambda}$ \citep{Kaspi+:2000,Bentz+:2009,Bentz+:2013}, as predicted from simple photoionisation physics \citep{Davidson:1972,Krolik+McKee:1978}. This strong correlation is the basis of single-epoch virial black hole mass estimators (`virial BH mass estimators' for short), which estimate the mass of the black hole using a single epoch of spectroscopy \citep[e.g.][]{Laor:1998,Wandel+:1999,McLure+Jarvis:2002,Vestergaard+Peterson:2006}. The single epoch mass estimation method is routinely used to estimate black hole masses \citep[e.g.][]{Vestergaard+:2008,Vestergaard+Osmer:2009,Willott+:2010,Schulze+Wisotzki:2010,Mortlock+:2011,Trump+:2011,Shen+Liu:2012,Kelly+Shen:2013}, and allows the black hole--galaxy correlations to be studied with large samples of galaxies. Several different virial BH mass estimators have been developed in the last decade, based on different line width characterisations and different lines. However, the FWHM of the H$\beta$, \ion{Mg}{ii}, and \ion{C}{iv} emission lines and a set value of $f$, are commonly used. Due to the widespread use of virial BH mass estimators, it is critical to fully understand the variation expected in $f$ within the AGN population.

The value of the $f$ factor depends on the structure, kinematics, dynamics, and orientation of the BLR with respect to the observer. Its value is expected to be different for every AGN. Nevertheless, it is a common practice to adopt a single $f$ factor value for all AGNs, calibrated from the local RM sample under the assumption that the $\subtext{M}{BH} \text{--} \sigma_{*}$ relation is consistent between quiescent and active galaxies \citep{Gebhardt+:2000,Ferrarese+:2001}. The value of $f$ also depends on line width characterisation (e.g.\ FWHM or \subtext{\sigma}{line}) and whether the mean or rms spectrum is used for the line width measurement. Recent measurements of $\langle f \rangle$ based on rms spectra and \subtext{\sigma}{line} vary between $\langle f_{\sigma} \rangle=2.8 \substack{+0.7 \\ -0.5}$ \citep{Graham+:2011} and $\langle f_{\sigma} \rangle=5.5 \pm 1.8$ \citep{Onken+:2004}, with most $\langle f_{\sigma} \rangle$ values lying within the range of $\langle f_{\sigma} \rangle \sim 4 \text{--} 6$ \citep{Collin+:2006,Park+:2012,Grier:2013,Woo+:2013,Pancoast+:2014,Woo+:2015}. The $\langle f_{\sigma} \rangle$ obtained from mean spectra is $3.85 \pm 1.15$ \citep{Collin+:2006}. On the other hand, the mean \subtext{f}{FWHM} measured using rms spectra were found to be $\langle \subtext{f}{FWHM} \rangle=1.12\substack{+0.36 \\ -0.27}$ by \citet{Woo+:2015} and $\langle \subtext{f}{FWHM} \rangle=1.44 \pm 0.49$ by \citet{Collin+:2006}. The $\langle \subtext{f}{FWHM} \rangle$ using mean spectra is $1.17 \pm 0.50$ \citep{Collin+:2006}. The calibration of the $f$ factor makes RM a secondary mass estimation method. The typical uncertainties in reverberation masses resulting from the uncertainty in $f$ is $\sim 0.43$ dex \citep{Woo+:2010}, due to the intrinsic scatter in the $\subtext{M}{BH} \text{--} \sigma_{*}$ relation.

Disk wind models provide a promising explanation for the observed broad absorption lines (BALs) and the blueshift of high-ionisation line relative to low-ionisation emission line, and are therefore the favoured model of the BLR (\citealt[][hereafter \citetalias{Murray+:1995}]{Murray+:1995}; \citealt[][hereafter \citetalias{Elvis:2004}]{Elvis:2004}). There is some evidence that for low ionisation lines such as H$\beta$, that the kinematics may be dominated by simple virialised rotation models \citep{Peterson+Wandel:1999,Kollatschny:2003,Kollatschny+Zetzl:Mar2013}. Thus, it is still an open question as to whether the disk wind models apply to some or all emission lines. We attempt to recover a theoretical prediction of the $f$ factor based on a dynamical disk wind model of the BLR and investigate the impact of orientation on the value of $f$.

The overview of the paper is as follows. In Section~\ref{sec:diskwind}, we describe our approach in modelling the disk wind. The results of the simulations are presented in Section~\ref{sec:results}. In Section~\ref{sec:discussion}, we discuss our findings and compare them to previous studies. The conclusion is summarised in Section~\ref{sec:summary}.

\section{Disk Wind Model} \label{sec:diskwind}

\begin{figure}
  \centering
  \includegraphics[width=\columnwidth]{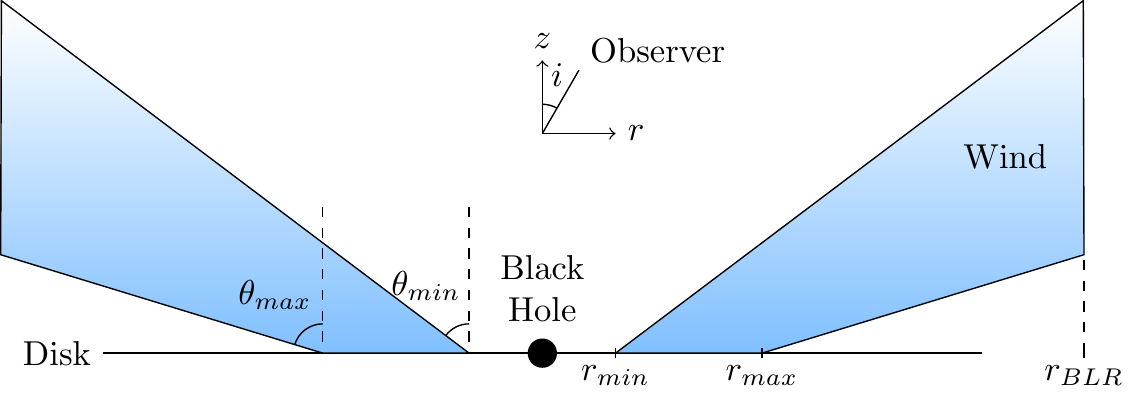}
  \caption{A sketch of the key parameters used to describe the cylindrical disk wind model.}
  \label{fig:cydiskmodel}
\end{figure}

Our BLR disk wind model is based on the cylindrical disk wind model introduced by \citet{Shlosman+Vitello:1993}, originally used to model cataclysmic variable stars (CVs). The model consists of a flat, opaque and geometrically thin accretion disk and a thick, conical BLR wind. A simplified sketch is shown in Figure~\ref{fig:cydiskmodel}. The similarities in the geometries, kinematics, and ionisation state between CVs and AGNs, suggest this model can be implemented to study the characteristics of AGN \citep{Higginbottom+:2013,Higginbottom+:2014}.

The properties of our cylindrical disk wind models are based on two of the well-known disk wind models, the \citetalias{Murray+:1995} line-driven disk wind model and the \citetalias{Elvis:2004} funnel disk wind model. The details of our model are given in the following sections.

\subsection{Kinematics of the Wind} \label{ssec:kinematics}

AGN are assumed to have an axially rather than a spherically symmetric geometry. Therefore, it is preferable to describe the model using cylindrical coordinates $(r, \phi, z)$. The variables $r$ and $\phi$ are the radial and azimuthal coordinates on the $xy$-plane, which is on the surface of the accretion disk. The rotation axis of the disk is aligned with the $z$ axis. The inclination angle, $i$, is defined from the $z$ axis to the observer's line of sight.

The outflowing wind is launched from the accretion disk at radii between \subtext{r}{min} and \subtext{r}{max} along fixed streamlines. Each point in the wind spirals upwards in three-dimensional helices with a fixed opening angle, $\theta$. The opening angle of each stream line is situated within a minimum angle, \subtext{\theta}{min}, and maximum angle, \subtext{\theta}{max}, and its value depends on the origin position of the streamline from the accretion disk, such that
\begin{equation}
\theta=\subtext{\theta}{min}+(\subtext{\theta}{max}-\subtext{\theta}{min})x^{\gamma},
\label{eqn:openingangle}
\end{equation}
where $x=(r_{0}-\subtext{r}{min})/(\subtext{r}{max}-\subtext{r}{min})$, $r_{0}$ is the origin position of a single streamline, and $\gamma$ is used to adjust the concentration of the streamlines toward either the inner or outer boundaries of the wind. Throughout our investigations we have set $\gamma=1$, which corresponds to even angular spacing between the streamlines.

The velocity components at any given position in the wind can be given in terms of the radial, rotational and vertical velocity, $v_{r}$, $v_{\phi}$, and $v_{z}$. Alternatively, the velocity can be expressed in terms of poloidal velocity, $v_{l}$, and rotational velocity, $v_{\phi}$. The poloidal velocity or the velocity along the streamline is
\begin{equation}
v_{l}=v_{0}+(v_{\infty}-v_{0})\left[\frac{(l/R_{v})^{\alpha}}{(l/R_{v})^{\alpha}+1}\right],
\label{eqn:poloidalvelocity}
\end{equation}
where $v_{0}$ is the initial poloidal wind velocity at the surface of the disk \citep[set arbitrarily at 6\,\kms;][]{Higginbottom+:2013,Shlosman+Vitello:1993}, $l=[(r-r_{0})^{2}+z^{2}]^{1/2}$ is the distance along a poloidal streamline, $R_{v}$ is the wind acceleration scale height characterising the scale at which the wind reaches half its terminal velocity $v_{\infty}$, and $\alpha$ is a power-law index that controls the shape of the acceleration profile. We set $\alpha=1$, which implies a slow increase in acceleration along each poloidal streamline. Furthermore, $v_{l}$ is correlated to $v_{r}$ and $v_{z}$ such that $v_{r}=v_{l}\sin\theta$ and $v_{z}=v_{l}\cos\theta$. In our model, the asymptotic wind velocity or the terminal velocity, $v_{\infty}$, is taken to be equal to the escape velocity, $\subtext{v}{esc}=(2G\subtext{M}{BH}/r_{0})^{1/2}$.

At the base of the wind, the rotational velocity is assumed to follow Keplerian motion, $v_{\phi,0}=(G\subtext{M}{BH}/r_{0})^{1/2}$. As the wind rises above the disk and expands, we assume the rotational velocity decreases linearly. This conserves angular momentum about the rotation axis
\begin{equation}
v_{\phi}=v_{\phi,0}\left(\frac{r_{0}}{r}\right).
\label{eqn:rotvelocity}
\end{equation}

For each position ($r,z$), the density of the wind, $\rho$, follows the continuity equation specified by
\begin{equation}
\rho(r,z)=\frac{r_{0}}{r}\frac{\diff r_{0}}{\diff r}\frac{\dot{m}(r_{0})}{v_{z}(r,z)},
\label{eqn:density}
\end{equation}
where the factor $(r_{0}/r)(\diff r_{0}/\diff r)$ scales as the streamline area increases with outflowing wind. The mass-loss rate per unit surface of the disk, $\dot{m}$, is
\begin{equation}
\dot{m}(r_{0})=\subtext{\dot{M}}{wind}\frac{r^{\lambda}_{0}\cos\theta(r_{0})}{\int \diff A\,r'^{\lambda}_{0}\cos\theta(r'_{0})},
\label{eqn:massloss}
\end{equation}
where \subtext{\dot{M}}{wind} is the total mass-loss rate of the wind, $\lambda$ is the mass-loss rate exponential, and the term $\cos\theta$ represents the angle formed between the streamline and the disk. A uniform mass-loss with radius is indicated by $\lambda=0$. For a high luminosity source $L \approx 10^{46}\,\ergs$ and black hole of mass $10^{8}\,M_{\odot}$, the total mass accretion rate is $\subtext{\dot{M}}{acc} \approx 2\,M_{\odot}\,$yr$^{-1}$ with efficiency $\eta=0.1$ \citep{Peterson:1997}. Here, \subtext{\dot{M}}{wind} is taken to be equivalent to \subtext{\dot{M}}{acc}.

\begin{figure}
  \centering
  \includegraphics[width=\columnwidth]{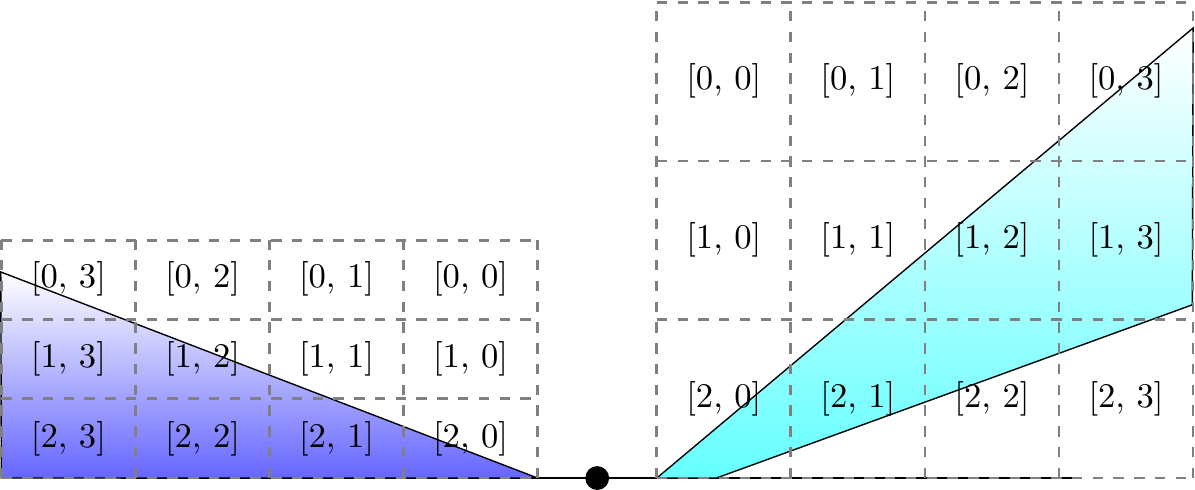}
  \caption{The numbered regions describe different `wind zones' where emission lines might be generated for \citetalias{Murray+:1995} ({\em left}, blue) and \citetalias{Elvis:2004} ({\em right}, cyan) disk wind models.}
  \label{fig:cydiskmodelzone}
\end{figure}

\begin{table*}
\caption{Adopted fiducial values of the parameters.}
\label{tab:parametersmodel}
\begin{center}
  \begin{tabular*}{\textwidth}{@{}l\x l\x c\x c@{}}
  \hline\hline
  Parameter & Notation & \citetalias{Murray+:1995}$^{a}$ & \citetalias{Elvis:2004}$^{b}$ \\
  \hline
  Black hole mass & \subtext{M}{BH} ($10^{8}\,M_{\odot}$) & 1.0 & 1.0 \\
  Wind radius & \subtext{r}{min}; \subtext{r}{max} ($10^{16}\,$cm) & 1.0; 10.0 & 1.0; 2.0 \\
  Wind angle & \subtext{\theta}{min}; \subtext{\theta}{max} & $69.0\degree;89.0\degree$ & $50.0\degree;70.0\degree$ \\
  Concentration of streamline & $\gamma$ & 1.0 & 1.0 \\
  Initial poloidal velocity & $v_{0}$ (\kms) & 6.0 & 6.0 \\
  Scale height & $R_{v}$ ($10^{16}\,$cm) & 20.0 & 25.0 \\
  Power law index & $\alpha$ & 1.0 & 1.0 \\
  Mass-loss rate exponent & $\lambda$ & 0 & 0 \\
  Total mass-loss rate & \subtext{\dot{M}}{wind} ($M_{\odot}\,$yr$^{-1}$) & 2.0 & 2.0 \\
  \hline\hline
  \end{tabular*}
\end{center}
\tabnote{$^{a}$ Chosen values to mimic \citet{Murray+:1995} model.}
\tabnote{$^{b}$ Chosen values to mimic \citet{Elvis:2004} model.}
\end{table*}

The relevant parameter values chosen for the \citetalias{Murray+:1995} and \citetalias{Elvis:2004} disk wind models are shown in Table~\ref{tab:parametersmodel}. The black hole mass was set to $10^{8}\,M_{\odot}$ with corresponding values of wind radius from \citetalias{Murray+:1995} and \citetalias{Elvis:2004}. In both cases, the wind region is defined out to the radius of the BLR, $\subtext{r}{BLR} \sim 10^{17}\,$cm. The \citetalias{Murray+:1995} and \citetalias{Elvis:2004} models have different heights and are scaled accordingly by \subtext{r}{BLR} and the wind angle, \subtext{\theta}{min} and \subtext{\theta}{max}, as illustrated in Figure~\ref{fig:cydiskmodelzone}. The opening angle of the \citetalias{Murray+:1995} model is close to the base of the accretion disk with a wider range of wind radii, and hence the height is shallow compared to the \citetalias{Elvis:2004} model. To account for the possibility that $\sim$~20\,\% of QSOs have BAL feature \citep{Knigge+:2008}, the wind opening angle is specified to be within $20\degree$ for both models. For simplicity, the vertical wind that is initially lifted off the disk in \citetalias{Elvis:2004} model is omitted.

There is strong evidence that the gas in the BLR is stratified, with high ionisation lines situated closer to the ionising source than the low ionisation lines \citep{Peterson+Wandel:1999,Kollatschny:2003,Peterson:2014}. We divide our disk wind model into different `wind zones' to account for the stratification of the wind in the BLR region. The wind is divided evenly into zones of 3 rows and 4 columns, as depicted in Figure~\ref{fig:cydiskmodelzone}. Each zone $[a, b]$ is labelled according to its row and column position.

\subsection{Line Profile Creation} \label{ssec:lineprofile}

Once the disk wind model is established, the expected emission line  profile is computed using a Monte Carlo simulation. Initially, a large number of particles dictated by the density profile, are randomly generated in cylindrical coordinates $(r, \phi, z)$ within the confines of the allowed `wind zone'. The projected velocity along the line-of-sight, \subtext{v}{los}, is then evaluated for each particles as a function of inclination angle, $i$. From the calculated line-of-sight velocity, kernel density estimation (KDE) is performed to estimate the shape of the underlying line profile. Line profiles are created for individual zones for inclination angles between $5\degree$ and $85\degree$. We assume that there is no obscuration or shielding due to the dusty torus. Photoioniation is not included in this model but will be incorporated in more detailed modelling to follow.

\subsection{The $f$ factor} \label{ssec:ffactor}

For each `wind zone' line profile, the FWHM and \subtext{\sigma}{line} values are measured and the corresponding $f$ factor is calculated using Equation~\ref{eqn:mbh}. As H$\beta$ is typically used to calculate the black hole mass in RM studies \citep{Collin+:2006,Park+:2012,Grier:2013,Woo+:2013,Pancoast+:2014,Woo+:2015}, we concentrate our analysis on a wind zone close to the base of the wind and towards the outer edge of the BLR, corresponding to expectations for H$\beta$ emission. To make the two models approximately comparable in terms of radial scales, zone [2, 2] is chosen for both models. We also calculate the probability of measuring a given black hole mass based on a fixed $f$ value from the literature using the cumulative probability of viewing a quasar at any given inclination angle, $i$, of $F(i)=1-\cos(i)$, with $0\degree \leq i \leq 90\degree$.

The response of an individual emission line to changes in the continuum flux is expected to vary depending on where the line is emitted within the disk wind and the luminosity of the AGN \citep[due to differences in density and ionising flux;][]{Korista+Goad:2000,Korista+Goad:2004}. Without further assumptions about the degree of continuum variability and luminosity and further photoionisation modelling, it becomes impossible to model rms spectra. Therefore, we only compare our generated spectra with the $f$ factors measured using the mean spectra.

\section{Results} \label{sec:results}

The range of $f$ factors found for the \citetalias{Murray+:1995} and \citetalias{Elvis:2004} disk wind models at various inclination angles and for various zones are presented in Table~\ref{tab:fzone}. The range of $f$ factors found extend well beyond the spread prescribed in the empirically determined $\langle f \rangle$ values \citep{Collin+:2006}.

\begin{table*}
\caption{Values of $f$ for different zones.}
\label{tab:fzone}
\begin{center}
  \begin{tabular}{@{\extracolsep{4pt}}l c c c c c c c@{}}
  \hline\hline
  Model & Zone & \multicolumn{3}{c}{\subtext{f}{FWHM}} & \multicolumn{3}{c}{$f_{\sigma}$} \\
  \cline{3-5}\cline{6-8}
  & & $i=5\degree$ & $i=45\degree$ & $i=85\degree$ & $i=5\degree$ & $i=45\degree$ & $i=85\degree$ \\
  \hline
  \citetalias{Murray+:1995} & [0, 3] & 16.22 & 0.34 & 0.17 & 132.91 & 3.09 & 1.57 \\
  & [1, 2] & 36.21 & 0.74 & 0.37 & 213.00 & 6.54 & 3.33 \\
  & [1, 3] & 32.76 & 0.63 & 0.31 & 175.46 & 5.56 & 2.83 \\
  & [2, 0] & 28.35 & 0.44 & 0.22 & 253.02 & 3.94 & 1.99 \\
  & [2, 1] & 28.13 & 0.45 & 0.22 & 248.33 & 4.13 & 2.09 \\
  & [2, 2] & 28.42 & 0.45 & 0.23$^{a}$ & 252.82 & 4.22 & 2.13$^{b}$ \\
  & [2, 3] & 28.57 & 0.45 & 0.23 & 251.72 & 4.12 & 2.08 \\
  \hline
  \citetalias{Elvis:2004} & [0, 3] & 14.75 & 0.37 & 0.19 & 95.03 & 3.38 & 1.73 \\
  & [1, 2] & 18.66 & 0.73 & 0.38 & 104.29 & 6.66 & 3.44 \\
  & [1, 3] & 13.15 & 0.54 & 0.28 & 77.26 & 4.90 & 2.53 \\
  & [2, 0] & 26.63 & 0.40 & 0.20 & 249.60 & 4.10 & 2.07 \\
  & [2, 1] & 41.40 & 0.89 & 0.44 & 221.76 & 7.90 & 4.02 \\
  & [2, 2] & 42.54 & 0.89 & 0.45$^{a}$ & 317.75 & 8.05 & 4.08$^{b}$ \\
  & [2, 3] & 40.72 & 0.71 & 0.36 & 394.98 & 6.46 & 3.26 \\
  \hline\hline
  \end{tabular}
\end{center}
\tabnote{\hspace{2cm} $^{a}$ Compare with $\langle \subtext{f}{FWHM(H$\beta$)} \rangle=1.17 \pm 0.50$ (or $\log_{10}\langle \subtext{f}{FWHM(H$\beta$)} \rangle=0.07 \substack{+0.15 \\ -0.24}$) from \citet{Collin+:2006}.}
\tabnote{\hspace{2cm} $^{b}$ Compare with $\langle f_{\sigma(\text{H}\beta)} \rangle=3.85 \pm 1.15$ (or $\log_{10}\langle f_{\sigma(\text{H}\beta)} \rangle=0.59 \substack{+0.11 \\ -0.15}$) from \citet{Collin+:2006}.}
\end{table*}

\begin{figure}
  \centering
  \includegraphics[width=0.98\columnwidth]{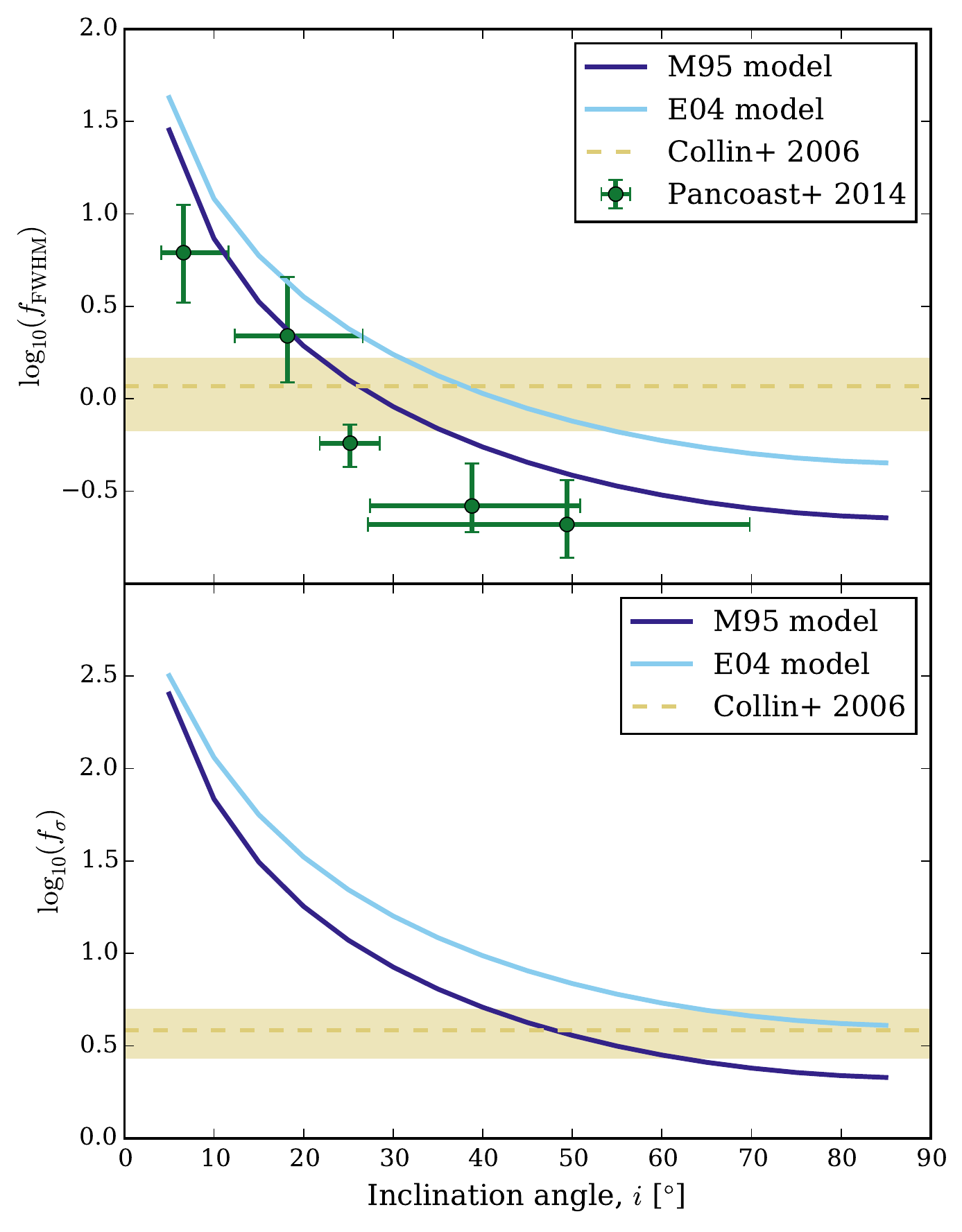}
  \caption{Plot of $f$ factors against inclination angle for H$\beta$ line characterised by emission from the [2, 2] location in the wind zone. {\em Upper}: Virial factor using FWHM, \subtext{f}{FWHM}. {\em Lower}: Virial factor using \subtext{\sigma}{line}, $f_{\sigma}$. The dashed lines are the mean $f$ factor, $\langle f \rangle$, with uncertainties (shaded) from \citet{Collin+:2006}. The \subtext{f}{FWHM(H$\beta$)} for individual quasars from \citet{Pancoast+:2014} are shown by green circles with error bars in the {\em upper} panel.}
  \label{fig:logfvsiZ22_ME}
\end{figure}

The distribution of $f$ factors with inclination angle for the equivalent H$\beta$ wind zone, in both models, is shown in Figure~\ref{fig:logfvsiZ22_ME}. The agreement between the empirically determined $f$ value from \citet{Collin+:2006} and our predictions varies between the two velocity dispersion characterisations. The \citet{Collin+:2006} \subtext{f}{FWHM} measurement coincided with middle range viewing angle (around $25\degree$ to $40\degree$) predictions using our disk wind models. Meanwhile, the \citet{Collin+:2006} $f_{\sigma}$ measurement was consistent with our prediction for a high inclination angle (edge-on) disk wind model.

Our predictions for $f$ were found to cover a similar range of values as those found using direct modeling estimates of \citet{Pancoast+:2014}. \citet{Pancoast+:2014} estimated the $f$ factor via direct BLR modelling using RM data of five Seyfert galaxies. Our predicted \subtext{f}{FWHM} values also follow the general trend with inclination of the \citet{Pancoast+:2014} results; however, our results display a systematic shift towards larger $f$ values.

\begin{figure}
  \centering
  \includegraphics[width=0.98\columnwidth]{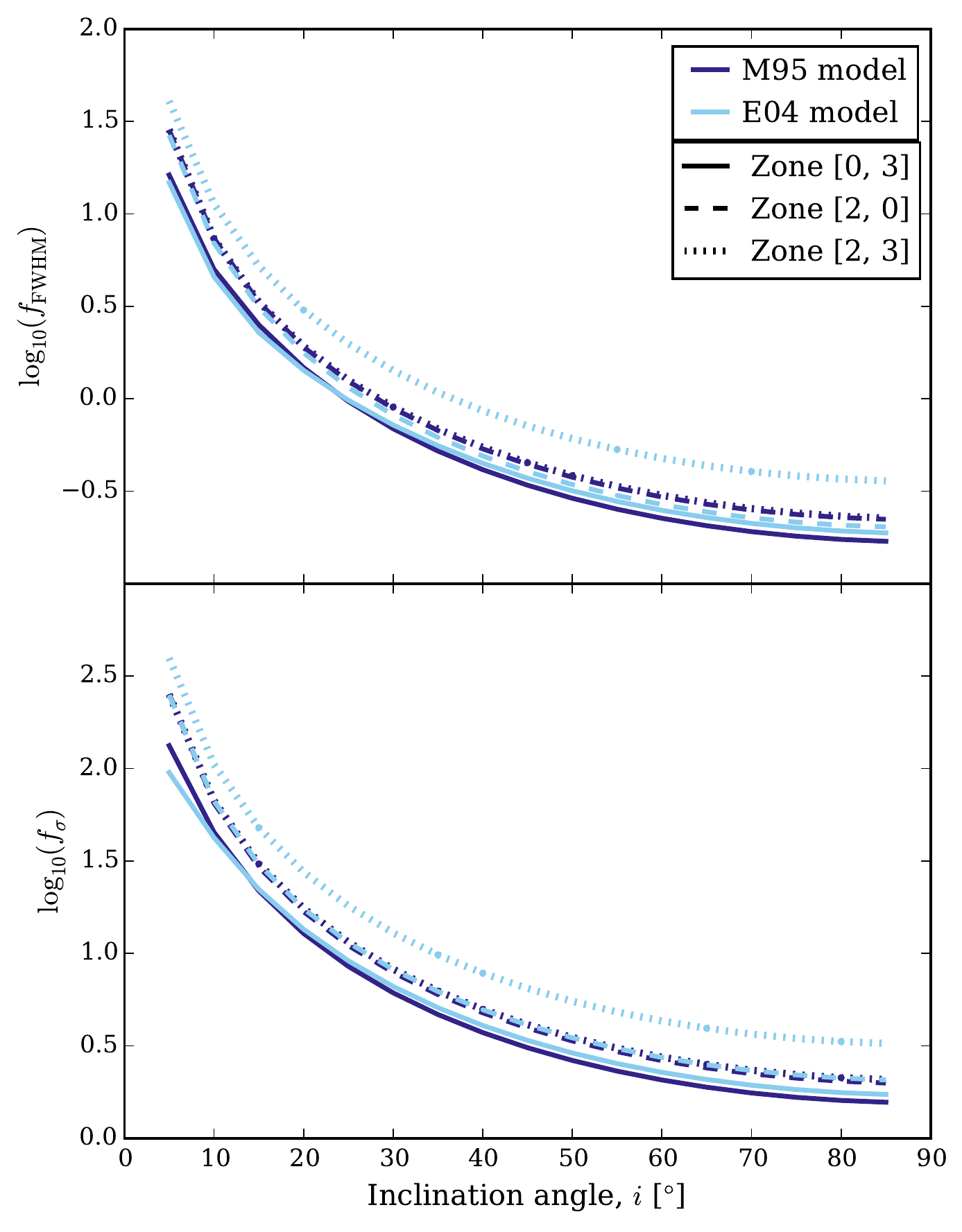}
  \caption{Plot of $f$ factors against inclination angle for wind zones [0, 3] (solid), [2, 0] (dashed), and [2, 3] (dotted) for the \citetalias{Murray+:1995} (blue) and \citetalias{Elvis:2004} (cyan) disk wind models.}
  \label{fig:logfvsiz_ME}
\end{figure}

The $f$ value as a function of inclination angle for selected wind zones [0, 3], [2, 0], and [2, 3] is illustrated in Figure~\ref{fig:logfvsiz_ME}. The zones provide some indication of the $f$ values for different emission lines expected to be emitted from different locations in the wind. The recovered values of $f$ are generally consistent between all wind zones. The wind is dominated by virialised rotational dynamics for wind zones close to the base of the wind. For wind zones at large $r$ and small $z$ (e.g.\ [2, 3]), $f$ has a steeper trend with inclination and its value is generally larger than the $f$ factor obtained in wind zones closer to the ionisation source (e.g.\ [2, 0]). In the \citetalias{Elvis:2004} model, the wind in zones [2, 1], [2, 2], and [2, 3] are located above the accretion disk (Figure~\ref{fig:cydiskmodelzone}, {\em right}). Since the initial positions of the streamlines, $r_{0}$, are contained within zone [2, 0], the rotational velocity in these outer zones rapidly diverges from Keplerian motion and quickly becomes smaller with larger $r$, in accordance with the conservation of angular momentum (see Equation~\ref{eqn:rotvelocity}). This results in the much larger $f$ values found in these zones compared with zone [2, 0]. However, as the poloidal velocity gradually increases and becomes dominant with increasing poloidal distance (that is, large $r$ and $z$; e.g.\ zone [0, 3]), the line width broadens and the true value of $f$ decreases.

\begin{figure*}
  \centering
  \includegraphics[width=0.98\textwidth]{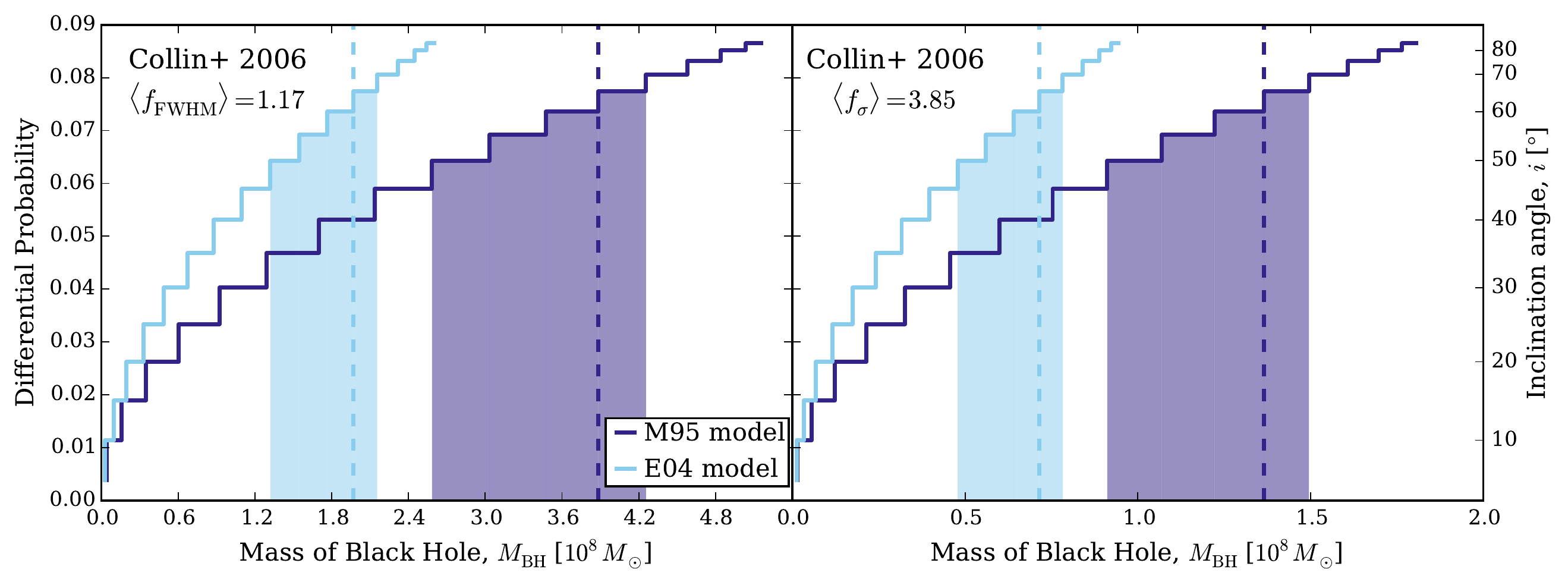}
  \caption{Differential probability against black hole mass for the \citetalias{Murray+:1995} (blue) and \citetalias{Elvis:2004} (cyan) models using $\langle f \rangle$ values for mean spectrum from \citet{Collin+:2006}. In these models, the true black hole mass is $10^{8}\,M_{\odot}$. The shaded region represents the \subtext{M}{BH} within one sigma range of the median (dashed). {\em Left}: Mean $f$ factor using FWHM of $\langle \subtext{f}{FWHM(H$\beta$)} \rangle=1.17$. {\em Right}: Mean $f$ factor using \subtext{\sigma}{line} of $\langle f_{\sigma(\text{H}\beta)} \rangle=3.85$.}
  \label{fig:diffprobvsmbhZ22_ME}
\end{figure*}

If a fixed value of $f$ is assumed, and the potential bias of orientation is ignored, then a large sample of quasars of the same mass will produce a broad distribution of black holes masses. To quantify the effect of the orientation dependence of $f$ on the black hole mass estimation for a large sample of AGN, we calculated the differential probability of estimating a given black hole mass using the fixed mean $f$ value from \citet{Collin+:2006}. The results are shown in Figure~\ref{fig:diffprobvsmbhZ22_ME}. Since the differential probability increases with increasing inclination angle, the possibility of seeing closer to edge-on is higher, $P(i)=\sin(i)$. Therefore, broader profiles are more likely to be observed as they become dominant in edge-on viewing angle, and black hole masses will be overestimated.

This is more pronounced in the FWHM case and in general, the mass determined from FWHM tends to be overestimated for both disk wind geometries. The median black hole mass recovered using the \citetalias{Murray+:1995} model was $3.88 \substack{+0.43 \\ -1.31} \times 10^{8}\,M_{\odot}$, approximately four times larger than the input black hole mass, and for the \citetalias{Elvis:2004} model, the median recovered black hole mass is double the input mass with $1.97 \substack{+0.22 \\ -0.66} \times 10^{8}\,M_{\odot}$.

The black hole masses obtained from \subtext{\sigma}{line} (Figure~\ref{fig:diffprobvsmbhZ22_ME}, {\em right}) tend to be less biased and more accurate in general. However, the accuracy of the recovered mass was still found to be model dependent. The median black hole mass recovered for the \citetalias{Murray+:1995} model of $1.37 \substack{+0.15 \\ -0.46} \times 10^{8}\,M_{\odot}$ is slightly overestimated, while the mass is underestimated in the \citetalias{Elvis:2004} model with $0.72 \substack{+0.08 \\ -0.24} \times 10^{8}\,M_{\odot}$.

\section{Discussion} \label{sec:discussion}

Obtaining accurate black hole mass measurements is crucial for understanding the role of black hole growth in galaxy evolution. Therefore, it is important to understand how the geometry and inclination of the BLR, and the chosen line width measurement affect the accuracy in our mass estimation.

The range of $f$ values can be much greater than the prescribed spread in the literature value. Therefore, we need to be cautious when using a single value of $f$ as it may bias mass estimates especially when the inclination angle is low. The \citetalias{Murray+:1995} and \citetalias{Elvis:2004} models establish relationships between the $f$ factor and inclination angle (Figure~\ref{fig:logfvsiZ22_ME}, {\em upper}), in agreement with \citet{Pancoast+:2014} despite different modelling approach.

The BLR geometry, kinematics and the origin of the emission line also affects the true $f$ value for individual AGN. This is evident from the differences in the $f$ values calculated for the different disk wind models, the offset between the disk wind models, the differences in the $f$ values for the different wind zones, and the dynamical modelling results of \citet{Pancoast+:2014}. The true nature of the BLR is unknown, and although some consistency is expected in the BLR, we currently cannot characterise the intrinsic distribution of $f$ values for the whole AGN population. However, the differences in $f$ due to the geometry and kinematics appears to be small compared to the effects of inclination angle. Also, our model assumes that the BLR is visible for all inclination angles. However, in the standard model of AGN, the BLR is believed to be obscured by a dusty torus. When this is taken into consideration, the estimate of the median black hole mass (Figure~\ref{fig:diffprobvsmbhZ22_ME}) is lowered as the probability of observing a closer to face-on AGN is increased.

Several studies have also investigated the scaling relationship between the $f$ factor and inclination angle \citep{Decarli+:2008,Kashi+:2013}. Our predicted \subtext{f}{FWHM} values are consistent with the analytical prediction of $f$ from \citet{Kashi+:2013} for a virialised line-driven disk wind model. However, the \citet{Decarli+:2008} predictions for a geometrically thin disk model better matches our findings for $f_{\sigma}$.

The line widths are typically measured using the FWHM or \subtext{\sigma}{line}. As the FWHM is a zeroth moment of the line profile, the sensitivity to the line core is higher than it is in the line wings. In contrast, \subtext{\sigma}{line} is a second moment of the line and is less affected by the line core. The \subtext{\sigma}{line} from rms spectra is commonly employed as a proxy in calculating the black hole mass since it has been argued that this provides a smaller bias and a better fit to the virial relation \citep{Peterson+:2004,Collin+:2006,Peterson:2011,Denney+:2013}. We found that the black hole mass estimated using the $\langle f_{\sigma(\text{H}\beta)} \rangle$ is closer to the input black hole mass compared to $\langle \subtext{f}{FWHM(H$\beta$)} \rangle$ value from \citet{Collin+:2006} (Figure~\ref{fig:diffprobvsmbhZ22_ME}). However, this discrepancy can be reduced when the shape corrections for $\langle \subtext{f}{FWHM(H$\beta$)} \rangle$ suggested by \citet{Collin+:2006}, are taken into consideration. When both $\langle \subtext{f}{FWHM(H$\beta$)} \rangle$ corrections \citep[Equation~5 and Equation~7;][]{Collin+:2006} are applied, the median black hole mass for the \citetalias{Murray+:1995} model is $1.66 \substack{+0.19 \\ -0.55} \times 10^{8}\,M_{\odot}$, which is marginally consistent with the true mass and the mass estimate found using $\langle f_{\sigma(\text{H}\beta)} \rangle$. For the \citetalias{Elvis:2004} model, the recovered median black hole masses are $0.84 \substack{+0.09 \\ -0.28} \times 10^{8}\,M_{\odot}$ after the $\langle \subtext{f}{FWHM(H$\beta$)} \rangle$ shape correction \citep[Equation~5;][]{Collin+:2006} and $0.94 \substack{+0.06 \\ -0.31} \times 10^{8}\,M_{\odot}$ using the FWHM width correction \citep[Equation~7;][]{Collin+:2006}.

It is worth mentioning several caveats in our modelling approach. The disk wind model simulations presented are simplifications of the complex BLR. The line driving mechanisms of the wind or photoionisation physics is not included in this model. We have also made major assumptions about the wind dynamics in our models, such as the local mass loss rate and the wind acceleration profile. The effects of these assumptions have not been investigated in this work. Future work will systematically search the parameter space in order to refine the models and to obtain a better fit with observations.

\section{Summary} \label{sec:summary}

In this work, we have implemented a dynamical disk wind prescription to explore the influence of BLR orientation on the black hole mass. The virial factor $f$, which scales the line-of-sight virial product to the true black hole mass, is calculated and compared to those from the literature. It is evident that the black hole masses recovered depend on several factors: the BLR geometry and dynamics, the origin of the emission line, and the inclination angle. The observed trend with inclination angle agrees with the results of \citet{Pancoast+:2014} despite the different models investigated. Additionally, the spread in the predicted $f$ values significantly exceeds the spread prescribed for the empirically determined values of $f$ from the literature. Therefore, using a single average value of $f$ may instill a bias into the mass estimate for large AGN samples.

We also computed the black hole mass using literature values of $f_{\sigma}$ and \subtext{f}{FWHM} from \citet{Collin+:2006}. The black hole mass is closer to the true mass if the velocity dispersion is measured using the \subtext{\sigma}{line}. Nevertheless, as suggested by \citet{Collin+:2006}, the \subtext{f}{FWHM} can be corrected to improve the mass estimate.

\begin{acknowledgements}
We thank the anonymous referee for valuable suggestions on the manuscript.
\end{acknowledgements}


\begin{thebibliography}{}
\expandafter\ifx\csname natexlab\endcsname\relax\def\natexlab#1{#1}\fi

\bibitem[{{Bentz} \& {Katz}(2015)}]{Bentz+Katz:2015}
{Bentz}, M.~C., \& {Katz}, S. 2015, PASP, 127, 67

\bibitem[{{Bentz} {et~al.}(2009){Bentz}, {Peterson}, {Netzer}, {Pogge}, \&
  {Vestergaard}}]{Bentz+:2009}
{Bentz}, M.~C., {Peterson}, B.~M., {Netzer}, H., {Pogge}, R.~W., \&
  {Vestergaard}, M. 2009, ApJ, 697, 160

\bibitem[{{Bentz} {et~al.}(2013){Bentz}, {Denney}, {Grier}, {Barth},
  {Peterson}, {Vestergaard}, {Bennert}, {Canalizo}, {De Rosa}, {Filippenko},
  {Gates}, {Greene}, {Li}, {Malkan}, {Pogge}, {Stern}, {Treu}, \&
  {Woo}}]{Bentz+:2013}
{Bentz}, M.~C., {Denney}, K.~D., {Grier}, C.~J., {et~al.} 2013, ApJ, 767, 149

\bibitem[{{Blandford} \& {McKee}(1982)}]{Blandford+McKee:1982}
{Blandford}, R.~D., \& {McKee}, C.~F. 1982, ApJ, 255, 419

\bibitem[{{Collin} {et~al.}(2006){Collin}, {Kawaguchi}, {Peterson}, \&
  {Vestergaard}}]{Collin+:2006}
{Collin}, S., {Kawaguchi}, T., {Peterson}, B.~M., \& {Vestergaard}, M. 2006,
  A\&A, 456, 75

\bibitem[{{Davidson}(1972)}]{Davidson:1972}
{Davidson}, K. 1972, ApJ, 171, 213

\bibitem[{{De Rosa} {et~al.}(2015){De Rosa}, {Peterson}, {Ely}, {Kriss},
  {Crenshaw}, {Horne}, {Korista}, {Netzer}, {Pogge}, {Ar{\'e}valo}, {Barth},
  {Bentz}, {Brandt}, {Breeveld}, {Brewer}, {Dalla Bont{\`a}}, {De
  Lorenzo-C{\'a}ceres}, {Denney}, {Dietrich}, {Edelson}, {Evans}, {Fausnaugh},
  {Gehrels}, {Gelbord}, {Goad}, {Grier}, {Grupe}, {Hall}, {Kaastra}, {Kelly},
  {Kennea}, {Kochanek}, {Lira}, {Mathur}, {McHardy}, {Nousek}, {Pancoast},
  {Papadakis}, {Pei}, {Schimoia}, {Siegel}, {Starkey}, {Treu}, {Uttley},
  {Vaughan}, {Vestergaard}, {Villforth}, {Yan}, {Young}, \&
  {Zu}}]{DeRosa+:2015}
{De Rosa}, G., {Peterson}, B.~M., {Ely}, J., {et~al.} 2015, ApJ, 806, 128

\bibitem[{{Decarli} {et~al.}(2008){Decarli}, {Labita}, {Treves}, \&
  {Falomo}}]{Decarli+:2008}
{Decarli}, R., {Labita}, M., {Treves}, A., \& {Falomo}, R. 2008, MNRAS, 387,
  1237

\bibitem[{{Denney} {et~al.}(2013){Denney}, {Pogge}, {Assef}, {Kochanek},
  {Peterson}, \& {Vestergaard}}]{Denney+:2013}
{Denney}, K.~D., {Pogge}, R.~W., {Assef}, R.~J., {et~al.} 2013, ApJ, 775, 60

\bibitem[{{Di Matteo} {et~al.}(2008){Di Matteo}, {Colberg}, {Springel},
  {Hernquist}, \& {Sijacki}}]{DiMatteo+:2008}
{Di Matteo}, T., {Colberg}, J., {Springel}, V., {Hernquist}, L., \& {Sijacki},
  D. 2008, ApJ, 676, 33

\bibitem[{{Di Matteo} {et~al.}(2005){Di Matteo}, {Springel}, \&
  {Hernquist}}]{DiMatteo+:2005}
{Di Matteo}, T., {Springel}, V., \& {Hernquist}, L. 2005, Nature, 433, 604

\bibitem[{{Du} {et~al.}(2015){Du}, {Hu}, {Lu}, {Huang}, {Cheng}, {Qiu}, {Li},
  {Zhang}, {Fan}, {Bai}, {Bian}, {Yuan}, {Kaspi}, {Ho}, {Netzer}, {Wang}, \&
  {SEAMBH Collaboration}}]{Du+:2015}
{Du}, P., {Hu}, C., {Lu}, K.-X., {et~al.} 2015, ApJ, 806, 22

\bibitem[{{Elvis}(2004)}]{Elvis:2004}
{Elvis}, M. 2004, in Astronomical Society of the Pacific Conference Series,
  Vol. 311, AGN Physics with the Sloan Digital Sky Survey, ed. G.~T. {Richards}
  \& P.~B. {Hall}, 109

\bibitem[{{Ferrarese} \& {Ford}(2005)}]{Ferrarese+Ford:2005}
{Ferrarese}, L., \& {Ford}, H. 2005, SSRv, 116, 523

\bibitem[{{Ferrarese} \& {Merritt}(2000)}]{Ferrarese+Merritt:2000}
{Ferrarese}, L., \& {Merritt}, D. 2000, ApJ, 539, L9

\bibitem[{{Ferrarese} {et~al.}(2001){Ferrarese}, {Pogge}, {Peterson},
  {Merritt}, {Wandel}, \& {Joseph}}]{Ferrarese+:2001}
{Ferrarese}, L., {Pogge}, R.~W., {Peterson}, B.~M., {et~al.} 2001, ApJ, 555,
  L79

\bibitem[{{Gebhardt} {et~al.}(2000){Gebhardt}, {Kormendy}, {Ho}, {Bender},
  {Bower}, {Dressler}, {Faber}, {Filippenko}, {Green}, {Grillmair}, {Lauer},
  {Magorrian}, {Pinkney}, {Richstone}, \& {Tremaine}}]{Gebhardt+:2000}
{Gebhardt}, K., {Kormendy}, J., {Ho}, L.~C., {et~al.} 2000, ApJ, 543, L5

\bibitem[{{Graham} {et~al.}(2001){Graham}, {Erwin}, {Caon}, \&
  {Trujillo}}]{Graham+:2001}
{Graham}, A.~W., {Erwin}, P., {Caon}, N., \& {Trujillo}, I. 2001, ApJ, 563, L11

\bibitem[{{Graham} {et~al.}(2011){Graham}, {Onken}, {Athanassoula}, \&
  {Combes}}]{Graham+:2011}
{Graham}, A.~W., {Onken}, C.~A., {Athanassoula}, E., \& {Combes}, F. 2011,
  MNRAS, 412, 2211

\bibitem[{{Grier}(2013)}]{Grier:2013}
{Grier}, C.~J. 2013, PhD thesis, The Ohio State University

\bibitem[{{G{\"u}ltekin} {et~al.}(2009){G{\"u}ltekin}, {Richstone}, {Gebhardt},
  {Lauer}, {Tremaine}, {Aller}, {Bender}, {Dressler}, {Faber}, {Filippenko},
  {Green}, {Ho}, {Kormendy}, {Magorrian}, {Pinkney}, \&
  {Siopis}}]{Gultekin+:2009}
{G{\"u}ltekin}, K., {Richstone}, D.~O., {Gebhardt}, K., {et~al.} 2009, ApJ,
  698, 198

\bibitem[{{Higginbottom} {et~al.}(2013){Higginbottom}, {Knigge}, {Long}, {Sim},
  \& {Matthews}}]{Higginbottom+:2013}
{Higginbottom}, N., {Knigge}, C., {Long}, K.~S., {Sim}, S.~A., \& {Matthews},
  J.~H. 2013, MNRAS, 436, 1390

\bibitem[{{Higginbottom} {et~al.}(2014){Higginbottom}, {Proga}, {Knigge},
  {Long}, {Matthews}, \& {Sim}}]{Higginbottom+:2014}
{Higginbottom}, N., {Proga}, D., {Knigge}, C., {et~al.} 2014, ApJ, 789, 19

\bibitem[{{Kashi} {et~al.}(2013){Kashi}, {Proga}, {Nagamine}, {Greene}, \&
  {Barth}}]{Kashi+:2013}
{Kashi}, A., {Proga}, D., {Nagamine}, K., {Greene}, J., \& {Barth}, A.~J. 2013,
  ApJ, 778, 50

\bibitem[{{Kaspi} {et~al.}(2000){Kaspi}, {Smith}, {Netzer}, {Maoz}, {Jannuzi},
  \& {Giveon}}]{Kaspi+:2000}
{Kaspi}, S., {Smith}, P.~S., {Netzer}, H., {et~al.} 2000, ApJ, 533, 631

\bibitem[{{Kelly} \& {Shen}(2013)}]{Kelly+Shen:2013}
{Kelly}, B.~C., \& {Shen}, Y. 2013, ApJ, 764, 45

\bibitem[{{King}(2003)}]{King:2003}
{King}, A. 2003, ApJ, 596, L27

\bibitem[{{King}(2005)}]{King:2005}
---. 2005, ApJ, 635, L121

\bibitem[{{Knigge} {et~al.}(2008){Knigge}, {Scaringi}, {Goad}, \&
  {Cottis}}]{Knigge+:2008}
{Knigge}, C., {Scaringi}, S., {Goad}, M.~R., \& {Cottis}, C.~E. 2008, MNRAS,
  386, 1426

\bibitem[{{Kollatschny}(2003)}]{Kollatschny:2003}
{Kollatschny}, W. 2003, A\&A, 407, 461

\bibitem[{{Kollatschny} \& {Zetzl}(2013)}]{Kollatschny+Zetzl:Mar2013}
{Kollatschny}, W., \& {Zetzl}, M. 2013, A\&A, 551, L6

\bibitem[{{Korista} \& {Goad}(2000)}]{Korista+Goad:2000}
{Korista}, K.~T., \& {Goad}, M.~R. 2000, ApJ, 536, 284

\bibitem[{{Korista} \& {Goad}(2004)}]{Korista+Goad:2004}
---. 2004, ApJ, 606, 749

\bibitem[{{Kormendy} \& {Ho}(2013)}]{Kormendy+Ho:2013}
{Kormendy}, J., \& {Ho}, L.~C. 2013, ARA\&A, 51, 511

\bibitem[{{Kormendy} \& {Richstone}(1995)}]{Kormendy+Richstone:1995}
{Kormendy}, J., \& {Richstone}, D. 1995, ARA\&A, 33, 581

\bibitem[{{Krolik} \& {McKee}(1978)}]{Krolik+McKee:1978}
{Krolik}, J.~H., \& {McKee}, C.~F. 1978, ApJS, 37, 459

\bibitem[{{Laor}(1998)}]{Laor:1998}
{Laor}, A. 1998, ApJ, 505, L83

\bibitem[{{Magorrian} {et~al.}(1998){Magorrian}, {Tremaine}, {Richstone},
  {Bender}, {Bower}, {Dressler}, {Faber}, {Gebhardt}, {Green}, {Grillmair},
  {Kormendy}, \& {Lauer}}]{Magorrian+:1998}
{Magorrian}, J., {Tremaine}, S., {Richstone}, D., {et~al.} 1998, AJ, 115, 2285

\bibitem[{{Marconi} \& {Hunt}(2003)}]{Marconi+Hunt:2003}
{Marconi}, A., \& {Hunt}, L.~K. 2003, ApJ, 589, L21

\bibitem[{{McConnell} \& {Ma}(2013)}]{McConnell+Ma:2013}
{McConnell}, N.~J., \& {Ma}, C.-P. 2013, ApJ, 764, 184

\bibitem[{{McLure} \& {Jarvis}(2002)}]{McLure+Jarvis:2002}
{McLure}, R.~J., \& {Jarvis}, M.~J. 2002, MNRAS, 337, 109

\bibitem[{{Mortlock} {et~al.}(2011){Mortlock}, {Warren}, {Venemans}, {Patel},
  {Hewett}, {McMahon}, {Simpson}, {Theuns}, {Gonz{\'a}les-Solares}, {Adamson},
  {Dye}, {Hambly}, {Hirst}, {Irwin}, {Kuiper}, {Lawrence}, \&
  {R{\"o}ttgering}}]{Mortlock+:2011}
{Mortlock}, D.~J., {Warren}, S.~J., {Venemans}, B.~P., {et~al.} 2011, Nature,
  474, 616

\bibitem[{{Murray} {et~al.}(1995){Murray}, {Chiang}, {Grossman}, \&
  {Voit}}]{Murray+:1995}
{Murray}, N., {Chiang}, J., {Grossman}, S.~A., \& {Voit}, G.~M. 1995, ApJ, 451,
  498

\bibitem[{{Murray} {et~al.}(2005){Murray}, {Quataert}, \&
  {Thompson}}]{Murray+:2005}
{Murray}, N., {Quataert}, E., \& {Thompson}, T.~A. 2005, ApJ, 618, 569

\bibitem[{{Onken} {et~al.}(2004){Onken}, {Ferrarese}, {Merritt}, {Peterson},
  {Pogge}, {Vestergaard}, \& {Wandel}}]{Onken+:2004}
{Onken}, C.~A., {Ferrarese}, L., {Merritt}, D., {et~al.} 2004, ApJ, 615, 645

\bibitem[{{Pancoast} {et~al.}(2014){Pancoast}, {Brewer}, {Treu}, {Park},
  {Barth}, {Bentz}, \& {Woo}}]{Pancoast+:2014}
{Pancoast}, A., {Brewer}, B.~J., {Treu}, T., {et~al.} 2014, MNRAS, 445, 3073

\bibitem[{{Park} {et~al.}(2012){Park}, {Kelly}, {Woo}, \& {Treu}}]{Park+:2012}
{Park}, D., {Kelly}, B.~C., {Woo}, J.-H., \& {Treu}, T. 2012, ApJS, 203, 6

\bibitem[{{Park} {et~al.}(2015){Park}, {Woo}, {Bennert}, {Treu}, {Auger}, \&
  {Malkan}}]{Park+:2015}
{Park}, D., {Woo}, J.-H., {Bennert}, V.~N., {et~al.} 2015, ApJ, 799, 164

\bibitem[{{Peterson}(1993)}]{Peterson:1993}
{Peterson}, B.~M. 1993, PASP, 105, 247

\bibitem[{{Peterson}(1997)}]{Peterson:1997}
---. 1997, {An Introduction to Active Galactic Nuclei}

\bibitem[{{Peterson}(2011)}]{Peterson:2011}
{Peterson}, B.~M. 2011, in Narrow-Line Seyfert 1 Galaxies and their Place in
  the Universe, 32

\bibitem[{{Peterson}(2014)}]{Peterson:2014}
---. 2014, SSRv, 183, 253

\bibitem[{{Peterson} \& {Wandel}(1999)}]{Peterson+Wandel:1999}
{Peterson}, B.~M., \& {Wandel}, A. 1999, ApJ, 521, L95

\bibitem[{{Peterson} {et~al.}(2004){Peterson}, {Ferrarese}, {Gilbert}, {Kaspi},
  {Malkan}, {Maoz}, {Merritt}, {Netzer}, {Onken}, {Pogge}, {Vestergaard}, \&
  {Wandel}}]{Peterson+:2004}
{Peterson}, B.~M., {Ferrarese}, L., {Gilbert}, K.~M., {et~al.} 2004, ApJ, 613,
  682

\bibitem[{{Peterson} {et~al.}(2014){Peterson}, {Grier}, {Horne}, {Pogge},
  {Bentz}, {De Rosa}, {Denney}, {Martini}, {Sergeev}, {Kaspi}, {Minezaki},
  {Zu}, {Kochanek}, {Siverd}, {Shappee}, {Araya Salvo}, {Beatty}, {Bird},
  {Bord}, {Borman}, {Che}, {Chen}, {Cohen}, {Dietrich}, {Doroshenko}, {Drake},
  {Efimov}, {Free}, {Ginsburg}, {Henderson}, {King}, {Koshida}, {Mogren},
  {Molina}, {Mosquera}, {Motohara}, {Nazarov}, {Okhmat}, {Pejcha}, {Rafter},
  {Shields}, {Skowron}, {Skowron}, {Valluri}, {van Saders}, \&
  {Yoshii}}]{Peterson+:2014}
{Peterson}, B.~M., {Grier}, C.~J., {Horne}, K., {et~al.} 2014, ApJ, 795, 149

\bibitem[{{Richstone}(1998)}]{Richstone:1995}
{Richstone}, D. 1998, in IAU Symposium, Vol. 184, The Central Regions of the
  Galaxy and Galaxies, ed. Y.~{Sofue}, 451

\bibitem[{{Schulze} \& {Wisotzki}(2010)}]{Schulze+Wisotzki:2010}
{Schulze}, A., \& {Wisotzki}, L. 2010, A\&A, 516, A87

\bibitem[{{Shen} \& {Liu}(2012)}]{Shen+Liu:2012}
{Shen}, Y., \& {Liu}, X. 2012, ApJ, 753, 125

\bibitem[{{Shlosman} \& {Vitello}(1993)}]{Shlosman+Vitello:1993}
{Shlosman}, I., \& {Vitello}, P. 1993, ApJ, 409, 372

\bibitem[{{Silk} \& {Rees}(1998)}]{Silk+Rees:1998}
{Silk}, J., \& {Rees}, M.~J. 1998, A\&A, 331, L1

\bibitem[{{Trump} {et~al.}(2011){Trump}, {Impey}, {Kelly}, {Civano}, {Gabor},
  {Diamond-Stanic}, {Merloni}, {Urry}, {Hao}, {Jahnke}, {Nagao}, {Taniguchi},
  {Koekemoer}, {Lanzuisi}, {Liu}, {Mainieri}, {Salvato}, \&
  {Scoville}}]{Trump+:2011}
{Trump}, J.~R., {Impey}, C.~D., {Kelly}, B.~C., {et~al.} 2011, ApJ, 733, 60

\bibitem[{{Vestergaard} {et~al.}(2008){Vestergaard}, {Fan}, {Tremonti},
  {Osmer}, \& {Richards}}]{Vestergaard+:2008}
{Vestergaard}, M., {Fan}, X., {Tremonti}, C.~A., {Osmer}, P.~S., \& {Richards},
  G.~T. 2008, ApJ, 674, L1

\bibitem[{{Vestergaard} \& {Osmer}(2009)}]{Vestergaard+Osmer:2009}
{Vestergaard}, M., \& {Osmer}, P.~S. 2009, ApJ, 699, 800

\bibitem[{{Vestergaard} \& {Peterson}(2006)}]{Vestergaard+Peterson:2006}
{Vestergaard}, M., \& {Peterson}, B.~M. 2006, ApJ, 641, 689

\bibitem[{{Wandel} {et~al.}(1999){Wandel}, {Peterson}, \&
  {Malkan}}]{Wandel+:1999}
{Wandel}, A., {Peterson}, B.~M., \& {Malkan}, M.~A. 1999, ApJ, 526, 579

\bibitem[{{Willott} {et~al.}(2010){Willott}, {Albert}, {Arzoumanian},
  {Bergeron}, {Crampton}, {Delorme}, {Hutchings}, {Omont}, {Reyl{\'e}}, \&
  {Schade}}]{Willott+:2010}
{Willott}, C.~J., {Albert}, L., {Arzoumanian}, D., {et~al.} 2010, AJ, 140, 546

\bibitem[{{Woo} {et~al.}(2013){Woo}, {Schulze}, {Park}, {Kang}, {Kim}, \&
  {Riechers}}]{Woo+:2013}
{Woo}, J.-H., {Schulze}, A., {Park}, D., {et~al.} 2013, ApJ, 772, 49

\bibitem[{{Woo} {et~al.}(2015){Woo}, {Yoon}, {Park}, {Park}, \&
  {Kim}}]{Woo+:2015}
{Woo}, J.-H., {Yoon}, Y., {Park}, S., {Park}, D., \& {Kim}, S.~C. 2015, ApJ,
  801, 38

\bibitem[{{Woo} {et~al.}(2010){Woo}, {Treu}, {Barth}, {Wright}, {Walsh},
  {Bentz}, {Martini}, {Bennert}, {Canalizo}, {Filippenko}, {Gates}, {Greene},
  {Li}, {Malkan}, {Stern}, \& {Minezaki}}]{Woo+:2010}
{Woo}, J.-H., {Treu}, T., {Barth}, A.~J., {et~al.} 2010, ApJ, 716, 269

\end{thebibliography}

\end{document}